\documentstyle{l-aa}
\input epsf

\def\teff{$T_{\rm eff}$}

\def\sun{$_\odot$}

\def\c2h2{C$_2$H$_2$}
\begin{document}
\thesaurus{ 08.01.3 - 08.03.1 - 08.01.1 - 13.09.6 }
\title{Spectral features of presolar diamonds in the laboratory and in 
carbon star atmospheres} 
\author{Anja C.\,Andersen  \inst{1} 
\and Uffe Gr{\aa}e J{\o}rgensen \inst{1} 
\and Flemming M.\,Nicolaisen \inst{2} 
\and Preben Gr{\aa}e S{\o}rensen \inst{2} 
\and Kristian Glejb{\o}l \inst{3}, \inst{4} }

\institute{Astronomisk Observatorium, Juliane Maries Vej 30, 
          DK--2100 Copenhagen, Denmark
\and
Institute of Chemistry, Universitetsparken 5, DK--2100 Copenhagen, Denmark
\and
Physics Institute, Technical University of Denmark, DK--2800 Lyngby, Denmark
\and
NKT Research Center A/S, Sognevej 11, DK--2605 Br{\o}ndby, Denmark}
\offprints{Anja C.\,Andersen, email address: anja@astro.ku.dk}
\date{Received date; accepted date}
\maketitle
\markboth{Spectral features of nano-diamonds}{A.\,C.\,Andersen et al.}

\begin{abstract}
Laboratory analyses on fine-grained diamond residues from primitive meteorites
have shown that nano-diamonds represent the most abundant form of presolar
dust preserved in meteoritic samples.  The presolar diamonds carry isotopic
anomalies which indicate a very complex formation history. Several 
groups of diamonds may exist with origin in different types of stars. In 
order to identify the sites of formation observationally,
we have extracted presolar diamonds from the Allende meteorite and 
measured the monochromatic absorption coefficient in a form which
is useful for stellar atmosphere calculations. The monochromatic absorption 
coefficient was measured in the wavelength ranges 400--4000 cm$^{-1}$
(2.5--25 $\mu$m) and 12200--52600 cm$^{-1}$ (190--820 nm).
We have made 
identical laboratory measurements on CVD diamonds as on the meteoritic
diamonds, in order to get a more solid basis for the interpretation of the 
diamond spectrum. 
The monochromatic absorption coefficient for the presolar diamonds was
incorporated in self-consistent carbon star photospheric models.
The main influence of the diamond dust in our photospheric models is a
heating of the upper photospheric layers and a
reduction of the C$_{2}$H$_{2}$ abundance. 
Due to the relatively small absorption coefficient of the diamonds
compared to other stellar dust grains, their spectral appearance is weak.
However, the weak interaction of the diamonds
with the radiation field may give them
an important role in the dust nucleation process.
The gas pressure will stay high and the gas
will be much closer to hydrostatic equilibrium
during possible diamond nucleation than is normally the case
in dust forming stellar regions, and therefore allow ample time for the
nucleation process.

\keywords{Stars: atmospheres - Stars: carbon - Stars:
abundances - Stars: infrared}

\end{abstract}

\section{Introduction}

Until 1987, dust particles around stars had only been recognised by their
appearance in stellar spectra, but the possibility of studying
unprocessed stellar condensates directly in the laboratory has given
important information not only about the Solar System formation, but has also
provided precise data for testing astrophysical stellar models.

Presolar diamonds were the first grains to be isolated 
from  meteorites of the carbonaceous chondrite type,  which could be
shown to have originated from outside the Solar
System (Lewis et al.\ 1987).  The identification of their presolar origin
has been possible on the basis of isotopic anomalies that appear inconsistent
with any known solar system process (Anders \& Zinner 1993).  
Primitive meteorites show the presence of two different components.  
One component consist of chondrules and other inclusions 
which have experienced various 
melting processes during the 
formation of the solar system.
The other component, called the matrix, is fine grained and have experienced
little or no heating.
It is in the matrix that the presolar grains
are found. The presolar diamond content is very
similar for all chondrite classes, around 500--1000 ppm of the matrix 
(Alexander et al.\ 1990; Huss 1990).                           
Their median grain size is about 2\,nm
(Fraundorf et al.\ 1989), which means that each diamond contain only about one 
thousand carbon atoms.  
A large fraction of the atoms are therefore on the surface of the crystal,
and
the meteoritic diamonds therefore consist of a mixture of diamond and
hydrogenate amorphous carbon (a-C:H).
The amorphous part has been estimated to account for 0.46 the volume fraction
of the presolar diamonds (Bernatowicz et al. 1990). 

The size distribution of the presolar diamonds
is log-normal rather than power-law,
reflecting growth rather than fragmentation and suggesting a short interstellar
residence time (Lewis et al.\ 1989). This size distribution is
surprising, as interstellar dust normally shows a power-law distribution,
which is the steady state form for fragmentation.
A log-normal distribution reflects 
either size-sorting or growth followed by partial conversion of small grains
to large ones (Lewis et
al.\ 1989). The good fit for the size distribution obtained by Lewis et al.\
(1989) down to small sizes suggests minimal contributions from processes that
affect small grains preferentially, such as fragmentation, sputtering and 
erosion. This suggests that the distribution is young and unevolved.  

Various mechanisms have been proposed to account for the production of
diamond grains in space, but the most likely scenario appears to be that the
nano-diamonds have condensed directly from stellar outflows (Lewis et al.\
1987; J{\o}rgensen 1988; Clayton 1989; Clayton et al.\ 1995).  
The conditions in cool stellar 
outflows are remarkably similar to those employed in industry to produce
diamonds by Chemical Vapour Deposition (CVD) (Angus \& Hayman 1988). 
The CVD mechanism makes use of the fact the the free energy difference
between diamond and graphite is only around 1 kcal/mol. Almost any chemical
reaction yielding graphite as the thermodynamically stable product can, in
principle, yield diamond as a metastable product. 
The trick is to steer the kinetics so as to favour diamond over graphite.
It has been suggested by Kr\"{u}ger et al.\ (1996), that in circumstellar
envelopes the surface growth processes on carbonaceous seed particles will
take place at sp$^{3}$ bonded carbon atoms rather than at sp$^{2}$ bonded
carbon atoms, which suggest that the grain material
formed in circumstellar envelopes will be amorphous diamond-like carbon.

For studies of radiative processes in stellar environments, knowledge of
opacities of the relevant atoms, molecules and grains are essential.
In order for the grains to be taken into account in stellar atmosphere
computations, knowledge of the spectral properties of the relevant grains are
needed.   Transmission spectroscopic 
measurements of presolar diamonds from the Allende meteorite 
in the ultra-violet, the optical and the infrared region 
are presented in this paper.
The measurements were designed such that it has been possible to determine 
the monochromatic absorption coefficient.  This coefficient is necessary in 
order to include the nano-diamonds in model atmosphere calculations and in
synthetic spectrum calculations. By use of the derived absorption coefficient,
synthetic spectra of carbon stars with the nano-diamonds included have 
been calculated, under the assumption that the diamonds we 
have extracted have the same optical properties as they had when they formed in
a stellar atmosphere.
Finally, we have performed
similar laboratory measurements on CVD diamonds, in
order to get a more solid basis for understanding
the spectral features of the presolar diamonds. 

\section{Experiments}

The presolar diamonds were extracted using
the method described by Tang \& Anders (1988).
A 10.8\,g piece of the Allende meteorite was dissolved by alternating treatment
with 10\,N HF--1\,N HCl and 6\,N HCl (10 ml/g meteorite) to remove 
the silicates. The sulfur was removed with a CS$_{2}$ treatment and the reactive
kerogen was destroyed by oxidation with HNO$_{3}$ and HClO$_{4}$.  
To separate the diamonds the sample was dispersed by ultrasonification in
0.1\,M NH$_{4}$OH, producing a diamond colloid, which was extracted after
centrifugation.

The CVD diamonds were prepared as a hetero-epitaxial diamond film on a
silicon (100) substrate which had been polished with 0.25 $\mu$g of diamond
powder before depositing the CVD diamonds.  The CVD diamonds were deposited
in a hot filament reactor with 8\% CH$_{4}$ in the H$_{2}$ source gas.  The
filament temperature was between 2200$^{\circ}$\,C and 2400$^{\circ}$\,C, the
substrate temperatures were between 700$^{\circ}$\,C and 900$^{\circ}$\,C and
the pressure was 5 mbar.  This gives a growth rate of about 1 $\mu$g/hour.
The CVD diamonds were scraped off from the silicon substrate with a diamond
needle.  

\begin{figure}
\vspace{4 cm}
\caption[]{
A transmission electron microscope bright-field micrograph showing a
cluster of presolar nm-sized diamond crystallite.  Each cluster consists of
about 1000 diamonds.  The corresponding diffraction pattern is shown 
indicating the (111), (220), (311) and (400) spacings for diamond crystallite
(d = 2.065 (2.06), 1.281 (1.26), 1.035 (1.0754) and 0.8892 (0.8916) {\AA},
respectively, with the table values in parenthesis).}
\end{figure}

Transmission electron microscopy (TEM) was carried out on 
both the presolar diamonds
and the CVD diamonds, 
using a Philips EM 430 transmission
electron microscope operated at 300 keV (Fig.\,1). With this instrument 
we performed conventional imaging, electron
diffraction and energy-dispersive X-ray spectroscopy (EDS) with a sensitivity
down to Boron. 
We estimated that the sizes of the CVD diamonds are between 1.4
nm and 14 nm, from visual inspection of the TEM images.  The CVD diamonds are
hence,  approximately three times bigger than the meteoritic diamonds.

The spectral measurements were carried out in the infrared (400--4000 
cm$^{-1}$; 2.5--25 $\mu$m) and in the UV/VIS (12200--52600 cm$^{-1}$; 190--820
nm) region.

The UV/VIS spectra of the presolar and CVD diamonds were taken on a very
dilute solution of 350 $\mu$g diamonds in distilled water.
The spectra were obtained with a HP 8452A Diode-Array
Spectrophotometer, which is a single-beam microprocessor-controlled 
spectrophotometer with a deuterium lamp as the light source giving 190 nm --
820 nm wavelength range.  The wavelength accuracy was $\pm$ 2 nm.  The
sample cell was a small quartz container with dimensions 40 mm $\times$
10 mm $\times$ 1 mm.  During the measurements, 
a reference spectrum of distilled water
was used.  The difference of the signals obtained for the quartz container
filled with distilled water and the one filled with diamonds in distilled water
was attributed to the absorption of the diamonds.

The infrared spectra of the presolar and CVD diamonds were obtained with
a FT-IR Perkin Elmer 1760 Infrared Spectrometer.  The spectrometer
operates in the region 400--5000 cm$^{-1}$.
The instrument operation is driven by a computer station which allowed
us to record the spectra in digital form.
All measurements were performed with 64 scans and a resolution of 1--2 
cm$^{-1}$.  
For sample preparation the KBr technique was used, where small quantities of
the sample are mixed throughly with powdered KBr (300 mg).  Because of the
softness of KBr and its bulk transparency between 40 and 0.2 $\mu$m, the KBr
and particle mixture can be pressed into a clear pellet (diameter = 13 mm)
(Colthup et al. 1990).  
Around 
300 $\mu$g of presolar diamonds was added to the KBr as a suspension 
in ethanol.  The diamonds were suspended in order to separate them,
so that they would not be stuck in big clumps, because the diamonds are 
likely to have formed individually, not 
in agglomeration in stellar atmospheres.
By suspending them we therefore expect to obtain a spectrum which 
compare best with the possible stellar spectrum. 
Around 200 $\mu$g of CVD diamonds
was added to the KBr in the same way and the KBr for the reference pellets were
also suspended in ethanol.  The ethanol was removed by pumping on the sample
before pressing the tablet.  Despite this, we still had traces of ethanol in
our spectra, so it was necessary to compensate for this by the use of the
reference spectra.

\section{Interpretation of the experimental spectra}

\subsection{The UV/VIS spectra of the diamonds}

The UV/VIS spectra of our  presolar and CVD diamond samples are very similar
to one another,
showing strong absorption in the UV, with 
a distinct band  at 217 nm 
(46\,000 cm$^{-1}$), a flat maximum around 270 nm (37\,000 cm$^{-1}$)
and declining  absorption towards the visual end of the spectrum (Fig.\,2). 
These absorption characteristics are also typical for terrestrial
diamond containing pairs of 
nitrogen atoms (Davies 1984).  
These bulk UV/VIS features deviate significantly from the results by Lewis et al. (1989),
but correspond well with the measurements obtained by Mutschke et al. (1995)
on presolar diamonds from the Murchison meteorite, except that the strength
of the flat maximum around 270 nm seem to be much stronger for the
Murchison presolar diamonds.

The position of the band around 217 nm coincides with that of the 
interstellar extinction bump at 217.5 nm, as it has also been found by 
Mutschke et al.\ (1995) for the Murchison meteorite.

\begin{figure}
\centering
\leavevmode
\epsfxsize=1.05
\columnwidth
\epsfbox{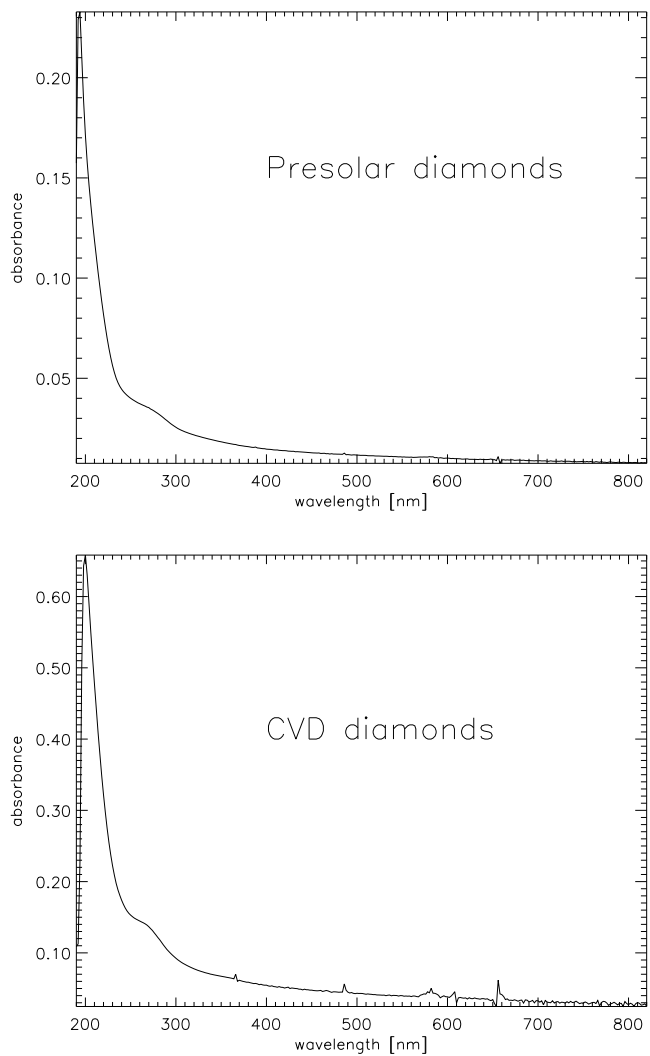}
\caption[]{The UV/VIS absorbance spectrum of the presolar and the CVD
diamonds, respectively.} 
\end{figure}

\subsection{The IR spectrum of the presolar diamonds}

The infrared absorbance spectrum of the presolar diamonds shows several
peaks (Table 1 and Fig. 3). 
The strong peaks are mainly due to impurities, meaning the presence of
atoms other than C. These impurities are most likely
situated at the grain surface, some of which may be due to the chemical
processing in the laboratory, other, of which may refer to the 
astrophysical environment the grains have experienced. 
The strong peaks at 3420 cm$^{-1}$  and 1632 cm$^{-1}$ 
are most likely due to H$_{2}$O (Chrenko et al. 1967).  
These peaks are strong and sharp, which makes tightly bound 
water a likely possibility (Nyquist \& Kagel 1971).  
This interpretation of the spectrum is further
supported by the peaks at 607 cm$^{-1}$ and 471 cm$^{-1}$.  

\begin{table}
\caption[]{Wavenumber of our measured IR bands, in cm$^{-1}$, 
detected in the spectra of the presolar
diamonds from the Allende meteorite and from CVD diamonds, respectively.
Also listed are our suggestions for assignments of the bands.}
\begin{flushleft}
\begin{tabular}{|l|l|l|}     \hline
Presolar     &       CVD       &       Assignment      \\ \hline
3420$^{sp}$  &    3418         &       O--H  stretch (in H$_{2}$O)  \\ \hline
3236$^{sh}$  &                 &       N--H stretch  \\  \hline
2954$^{sh}$  &    2972         &       C--H stretching from CH$_{3}$/CH$_{2}$ \\
2924         &    2927         &       groups  \\
2854         &    2856         &                         \\ \hline
             &    1719         &       C=O stretch (strained ring)  \\ \hline 
1632$^{sp}$  &    1640         &       O--H bend (in H$_{2}$O)    \\  \hline
1462         &                 &       C--H deformation (CH$_{3}$/CH$_{2}$) \\ 
1456         &    1456         &       \\ \hline
1402$^{sh}$  &    1401$^{sh}$  &       C--H deformation (CH$_{3}$)/inter-  \\
1385$^{sp}$  &    1385$^{sp}$  &       stitial N \\   \hline
1122         &    1122         &       C--O/C--N/C--C stretch     \\
1109         &    1112         &          \\ 
1090         &    1089         &                               \\
1054         &    1051         &                          \\ \hline
721$^{w}$    &                 &       C--H out-of-plane bend in \\
             &                 &       alkene residues   \\  \hline    
 633$^{sh}$  &                 &       C--Cl stretch     \\ \hline
 607         &     579$^{br}$  &       O--H    bend/torsion in  bonded  \\
 471         &                 &       water            \\ \hline
\end{tabular}
\end{flushleft}
\footnotesize{sp = sharp; sh = shoulder; br = broad; w = weak}   
\end{table}
\noindent

The shoulder peak at 3236 cm$^{-1}$ can be assigned to N--H stretching.

The peak frequencies and shapes of the triplet at
2954 cm$^{-1}$, 2924 cm$^{-1}$ and 2854 cm$^{-1}$, can be attributed to
long aliphatic hydrocarbon chains C$_{\rm n}$ with n$\ge$8
and 
indicating the presence of both
--CH$_{3}$ and --CH$_{2}$ symmetric and antisymmetric deformation modes.
This is further supported by part of 
the  broad peak between  1462 cm$^{-1}$ and 
1385 cm$^{-1}$ and the weak absorption at 
721 cm$^{-1}$ (Trombetta et al.\ 1991). The broad peak between  1462 cm$^{-1}$
and 1385 cm$^{-1}$ can also partly be assigned to interstitial N in 
diamond (Lewis et al. 1987). 

\begin{figure}
\centering
\leavevmode
\epsfxsize=1.05
\columnwidth
\epsfbox{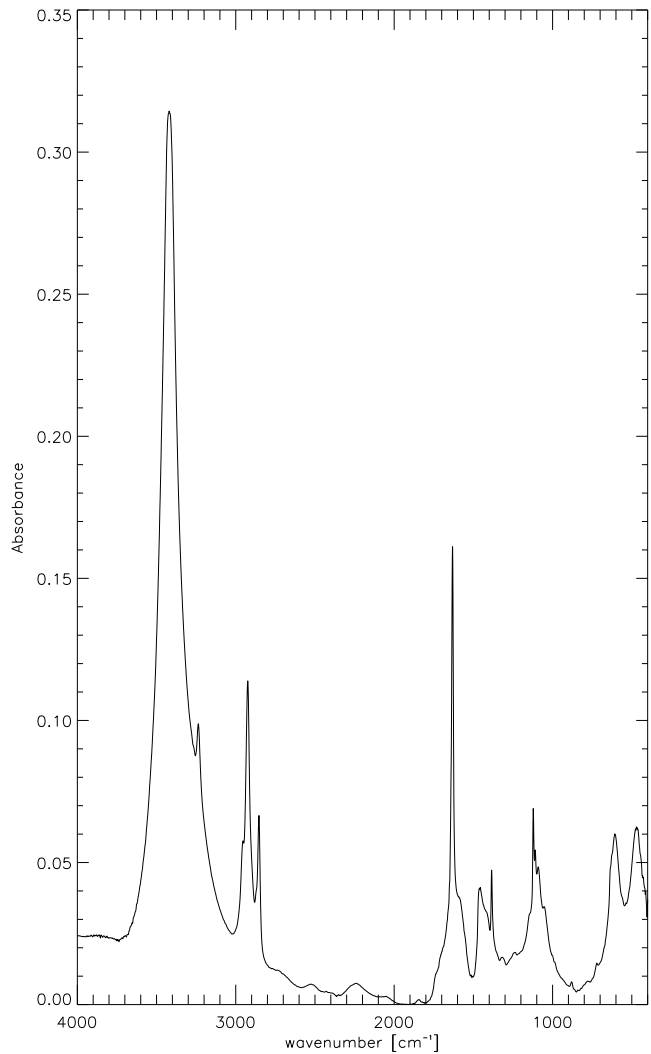}
\caption[]{Infrared absorbance spectrum of the 
presolar diamonds from the Allende meteorite.}
\end{figure}

The broad peak at 1122--1054 cm$^{-1}$ can be attributed to C--O stretching in
aliphatic ethers and/or C--N stretching.

The shoulder at 633 cm$^{-1}$ could be due to a C--Cl stretch, see later in
the text.

\begin{table*}
\caption[]{Spectral bands, in cm$^{-1}$ unless otherwise stated, 
detected in the obtained spectra of the presolar
diamonds from the Allende, Murchison and Orgueil meteorites. The different
interpretations by the various authors are indicated by corresponding numbers.}
\begin{flushleft}
\begin{tabular}{|l|l|l|l|l|l|l|}     \hline
\multicolumn{3}{|c|}{ALLENDE} & \multicolumn{2}{|c|}{MURCHISON} &
                ORGUEIL & Assignment by different authors \\ \hline
$^{1}$This & $^{2}$Lewis  &  $^{3}$Koike$^{ab}$ &  $^{4}$Colangeli &
$^{5}$Mutschke  & $^{6}$Hill &     \\
~paper & ~~et al. & ~~et al.  & ~~et al.  & ~~et al.  & ~~et al.  & \\
 & ~~1989 & ~~1995 &  ~~1994 & ~~1995 & ~~1996 & \\
\hline \hline
217 nm &  &  &    & 217 nm &  & $^{5}\pi-\pi^{*}$ transition of aromatically
 \\ 
270 nm &  & &  & 270 nm  & & bonded C, $^{1}$paired N in diamond \\ \hline
3420 & 3402 & 3401 & 3450 & 3400 &  &  $^{2}$O--H in --COOH,
$^{1,3,4,5,6}$O--H stretch (in H$_{2}$O)  \\ \hline
3236 & 3210 & 3115 &  3164 &  &  &   $^{1,2,3,4,5,6}$N--H stretch  \\  \hline
2954 & & 2976 &  & 3000 & 2985   & $^{1,2,3,4,5,6}$C--H stretching from CH$_{3}$/CH$_{2}$  \\
2924 & 2919 & 2924  & & & 2935    &  groups       \\ 
2854 & 2849 & 2849  & & 2800 & 2875  &         \\ \hline
& 1774 &  & 1771 & 1746 & 1728 &   $^{2,4}$C=O in --COOH, $^{1}$C=O (strained ring)   \\
     & & & & & &  $^{5}$C=O bonds in ester,  $^{6}$C=O or C=C \\ \hline
1632 & 1640 & 1634 & 1616 &  &  & $^{2,4}$aromatic C=C/C=O stretch,
   \\
& & 1590 & & & & $^{3}$C=O/N-H stretching, $^{1}$O--H bend (in H$_{2}$O) \\ \hline
1462 &  &   &  &    &     &  $^{1}$C--H deformation (CH$_{3}$/CH$_{2}$) \\
1456 &  &  & & & &      \\ \hline
1402 & 1403 & 1401 & & & & $^{1,2,4}$C--H deformation (CH$_{3}$)/interstitial N \\ 
1385 & 1361 & 1399  & 1399 & &  &  \\ \hline
 & 1234 &  &  &  & 1289  &  $^{3,4,5,6}$C--O/C--N stretch \\ 
 & 1173 & 1178 &  & 1175 &  & $^{1}$C--O/C--N/C--C stretch/interstitial N  \\
1122 &  & 1122 & &  & 1125 &  $^{2}$C--O stretch/interstitial N \\
1109  & 1108 & 1108 &  &  &  & \\
1090 & 1090  & 1080  & 1084 & 1090  & &   \\
1054 & 1028  &  &  &  & 1072 & \\ \hline
721 & &  & 744 & 725 & 745 & $^{1}$C--H out-of-plane bend in alkene residues \\ \hline
633 & 637 & 636 & & 630  & &  $^{1}$C--Cl stretch \\
& &630 & & & &  \\ 
& &626 & & & &  \\ \hline
607 & & & & &  &  $^{1}$O--H bend/torsion in water  \\
471 & & & & & &               \\ \hline
& & & & 396 & &  $^{5}$C=O=C or C=N=C \\
& & & & 367 & & \\
& & & & 310 & & \\ \hline
& & & & 130 & & ?? not yet identified  \\
& & & & 120 & & \\ \hline
\end{tabular}
\end{flushleft}
\footnotesize{$a$ = the spectra were obtained on diamond-like residues. \\
$b$ = transformed to cm$^{-1}$ from their tabulated values given in $\mu$m.}
\end{table*}
\noindent

\subsubsection{The O--H bands}

The indication of 
tightly bound water which may be intrinsic, is not found in the other spectra 
obtained on presolar diamonds from 
the Allende meteorite that has been published (Lewis et al. 1989; 
Koike et al. 1995). See Table 2.
Normally a diamond cannot be wetted with water, but under certain conditions
the water-repelling nature of the surface disappears (Davies 1984.)  
It has been shown for terrestrial diamonds that if the diamond is
heated to a few hundred degrees centigrade in O$_{2}$, 
the water-repelling nature
of the surface disappears; while heating the diamond in vacuum or in an
atmosphere of H restores its water-repelling nature (Sappok \& Boehm 1968).
This means that, if these water features are intrinsic (as oppose to being
an effect of not fully reduced water spectral features from humidity in
the KBr pellet), they
cannot originate from when the diamonds
were formed in a carbon/hydrogen 
rich atmosphere, but rather is an artifact which most
likely originates from the very rough chemical treatment that was used in
order to extract the diamonds.

The peak around 3400 cm$^{-1}$ obtained 
by Lewis et al. (1989), for diamonds from the Allende meteorite, 
was  interpreted by them
as coming from --COOH.  We consider this peak to be too 
high in wavenumber and too narrow to be assigned OH
in --COOH, while we find a good fit with known spectral bands of
water.  Also the band at
1774 cm$^{-1}$ is somewhat high in wavenumber for a --COOH interpretation.  
The --COOH interpretation has also been questioned by Mutschke et al.\ (1995)
and Hill et al.\ (1996).
Variations in the intensity
and in the presence of a peak around 1740 cm$^{-1}$ 
have also been observed in the
spectra of CVD diamonds, where the peak varied considerably for spectra
obtained at different times and with different IR spectrometers (Janssen 1991).

\subsubsection{The C-H bands}

The triplet around 2900 cm$^{-1}$ and the peaks around 1400 cm$^{-1}$ can
all be ascribed to the presence of  
saturated hydrocarbons (aliphatic compounds), but  
the peaks around 1400 cm$^{-1}$ are also 
in the right range for N in diamond (Lewis et al.\ 1989).  

\subsubsection{The nitrogen bands}

The N--H stretch at 3236 cm$^{-1}$ and the
broad band around 1100 cm$^{-1}$  
present in most of the published spectra of meteoritic
diamonds can be ascribed to various forms of
nitrogen in and on the diamonds. Presolar diamonds have been shown by
Russel et al. (1991) to be nitrogen rich.

Part of the 1100 cm$^{-1}$ peak can also be attributed to C--O stretching in
aliphatic ethers but the feature around 1084 cm$^{-1}$ could just as well
be due to C--N stretching and is in the right range for single N in terrestrial
diamonds (Clark et al.\ 1979).  

\subsubsection{The C--Cl bands}

We attribute the weak shoulder at 633 cm$^{-1}$ to C--Cl stretch based on
a comparison with the spectra obtained by Koike et al. (1995), of 
diamond-like residues
from the Allende meteorite and on the fact that terrestrial diamonds
does not show absorption in this wavelength region (Davies 1977).  
In the spectra obtained by Koike et al.
(1995) they have a very sharp peak at 630 cm$^{-1}$
which fits the pattern of a CCl stretch and the small splitting of the feature
in their spectra could be due to isotopic effects ($^{35}$Cl, $^{37}$Cl).  
Since they have used part of the same
treatment as we did, it seems very likely that both spectra could contain an
impurity which originates from HCl.

\subsection{The IR spectrum of the CVD diamonds}

The infrared absorbance spectrum of the CVD diamonds also show several peaks, 
many of which correspond to the peaks found for the presolar diamonds.  
The primary differences between the spectral properties of the presolar diamonds
and the CVD diamonds, are that the CVD diamonds have spectral 
characteristics closer related to soot than have the presolar diamonds.

\begin{figure}
\centering
\leavevmode
\epsfxsize=1.05
\columnwidth
\epsfbox{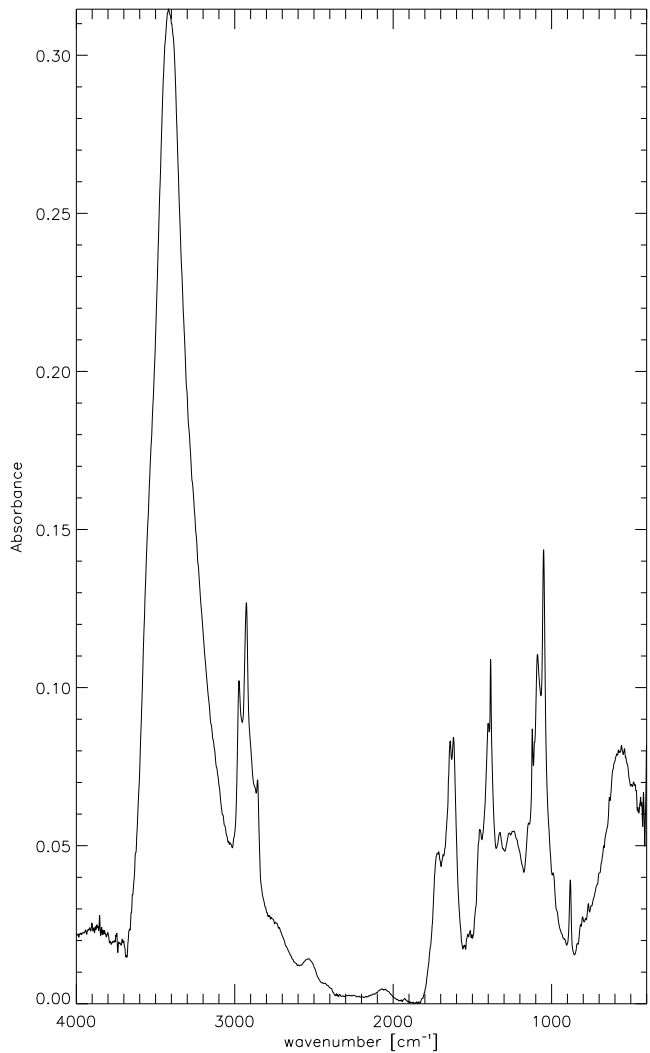}
\caption[]{Infrared absorbance spectrum of the CVD diamonds. 
Due to the compensation
for the ethanol in which the diamonds were suspended (see section 2),
the spectrum is over compensated around 3640 cm$^{-1}$ and 1640 cm$^{-1}$.}
\end{figure}

The nitrogen which is
present in the CVD diamonds is due to nitrogen impurities in the 
CH$_{4}$/H$_{2}$ gas mixture that was used during the deposition (Locher et al.
1994).

The peaks we attribute to O--H at 3418 cm$^{-1}$, 1640 cm$^{-1}$ 
and 579 cm$^{-1}$ 
are broader for the CVD diamonds
than for the presolar diamonds,  indicating that the water present here is not
tightly bound.

\subsection{Discussion}

The IR spectra of pure diamond have very few absorption features.
For terrestrial diamonds  the intrinsic absorption
present in all types of diamonds are at wavenumber greater than 1400 cm$^{-1}$
($\lambda < 7.1 \mu$m), with a peak around 2000 cm$^{-1}$ (5 $\mu$m)
for pure diamond, while that below 1400 cm$^{-1}$ ($\lambda > 7.1 \mu$m)
is specimen dependent both in strength and shape
(Davies 1977) and is caused by impurities, typically nitrogen.

The features that are present in the spectra of presolar diamonds are 
results of O--H, N--H, C--H, C=O, C=C, C--O, C--N and C--C interactions.
It is not possible by IR-spectroscopy to distinguish whether the
impurities in or on the diamonds are due to the chemical processing in the
laboratory, in the interstellar space, during the Solar System formation
or whether the "impurities" are original features which refer to the astrophysical 
environment the grains have experienced.  It has been shown by Russel et al. (1991)
that the presolar diamonds carry isotopic anomalous nitrogen with
$^{14}$N/$^{15}$N = 406 (terrestrial = 272) and by Virag et al. (1989) that the
presolar diamonds also carry anomalous hydrogen with $^{1}$H/$^{2}$D = 5193
(terrestrial = 6667) implying that at least some of the H and N must be
presolar.  From a spectroscopic point of view it does not really matter
whether the hydrogen bonds on the surface of the grains are ``original''
from the stellar environment or have been replaced at a later stage, as long
as the presence of hydrogen is to be expected on the grain surface in the
stellar environment.  Of course, then a replacement by a different element will
make a difference.

This makes the whole discussion about facts and artifacts in the spectra of
presolar diamonds complicated, since if the presolar diamonds are exposed to
a very rough chemical treatment to clean them as much as possible, it 
might be possible to  get a clean surface of the diamonds, but the obtained
spectra might not resemble what we can expect to observe 
if the nano-diamonds, at their
place of origin, are not clean but  
characterised by C--N, N--H and C--H bonds on the surface.

\section{Monochromatic absorption coefficient}

The monochromatic absorption coefficient $\kappa$
can be derived from the Beer-Lambert law,
\begin{equation}
\frac{I}{I_{0}} = e^{ - \rho \cdot l \cdot \kappa}
\label{beerlam}
\end{equation}
where $I_{0}$ and $I$ are the intensities of the incident and transmitted
light, respectively, $\rho$ is the density of the absorbing substance,
$l$ is the path length of the absorbing substance
and $\kappa$ is the monochromatic absorption coefficient.

For the UV/VIS measurements the absorption is measured  as
log$_{10}$(I$_{0}$/I), this give the following expression for the
monochromatic absorption coefficient after Eq.\,(\ref{beerlam}):
\begin{equation}
\kappa  =   \frac{1}{\rho \cdot l} \cdot {\rm ln} \frac{I_{0}}{I}
        =   \frac{2.30258}{\rho \cdot l} \cdot {\rm log_{10}} \frac{I_{0}}{I}.
\label{kappa}
\end{equation}
The bulk density $\rho$ of the presolar diamonds have been found by Lewis et al.
(1989) to be 2.22 -- 2.33 g/cm$^{3}$. The path length $l$ through the diamonds
can be found as the volume of diamond sample divided by the area of the
sample.  The area of the diamond sample is identical to the area
of the sample cell if the diamond sample is dispersed
evenly in the sample cell.
$l$ is then the effective path length of the radiation through an
equivalent, thin diamond film of the equivalent  thickness
$l = $V$_{\rm dia}$/A$_{\rm cell}$ = V$_{\rm dia}$/A$_{\rm dia}$ =
$\frac{m_{\rm dia}}{\rho_{\rm dia}}$/A$_{\rm dia}$, so that
$\rho_{\rm dia} \cdot l = $ m$_{\rm dia}$/A$_{\rm dia}$.
\begin{eqnarray}
\kappa_{\rm UV} & = & \frac{2.30258}{m_{\rm dia}/{\rm A}_{\rm dia}}{\rm log_{10}}
                     \cdot  \frac{I_{0}}{I}  \\
   & = & \frac{2.30258}{\left(350\times10^{-6}{\rm g}
\right)/((4.0{\rm cm})(1.0 {\rm cm}))} \cdot {\rm log_{10}}\frac{I_{0}}{I} \nonumber \\
   & = &  26315 \cdot {\rm log_{10}}\frac{I_{0}}{I} \hspace{0.4 cm} {\rm cm^{2}/ g} \nonumber.
\label{UV}
\end{eqnarray}

\begin{eqnarray}
\kappa_{\rm IR} & = & \frac{2.30258}{m_{\rm dia} / {\rm A}_{\rm dia}} \cdot {\rm log_{
10}} \frac{I_{0}}{I}  \\
   & = & \frac{2.30258}{\left(300 \times 10^{-6} {\rm g} \right) / (\pi \left( \frac{1.3}{2} {\rm cm} \right)^{2})} 
\cdot {\rm log_{10}}\frac{I_{0}}{I} \nonumber \\
    & = &  10188 \cdot {\rm log_{10}}\frac{I_{0}}{I} \hspace{0.2 cm} {\rm cm^{2} /g} \nonumber.
\label{IR}
\end{eqnarray}
The transmittance, T, is:
\begin{eqnarray*}
T = \frac{I}{I_{0}} \times 100\%,
\end{eqnarray*}
and the absorbance, D, is given by
\begin{eqnarray*}
D = {\rm log_{10}} \frac{I_{0}}{I}.
\end{eqnarray*}

Since the transmittance measurement give the monochromatic extinction
coefficient (extinction = absorption + scattering) it is necessary to
correct for scattering before the monochromatic absorption coefficient can
be derived.  The measured IR-spectra were corrected for scattering by
performing a base-line correction on the measured transmittance spectrum,
before it was converted into an absorbance spectrum.
In the base-line method  a base line is drawn tangent to the spectrum at the
wings of the analytical band, and its intersection with a vertical line at
the analytical wavenumber is used for $I_{0}$.
No correction for scattering was performed for the UV/VIS measurements, since
the amount of scattering in the experiment was very small.

The primary reason for scattering in the measured IR transmittance spectrum is
that when the sample is incorporated in a KBr matrix the size of the KBr
grains will force the sample grains to lay
around the much larger KBr
grains, resulting in some clumping of the sample.  The scattering is
therefore mainly a matrix effect, and not a property of the diamond that we
should expect to observe in stellar environments.

\begin{figure}
\centering
\leavevmode
\epsfxsize=1.05
\columnwidth
\epsfbox{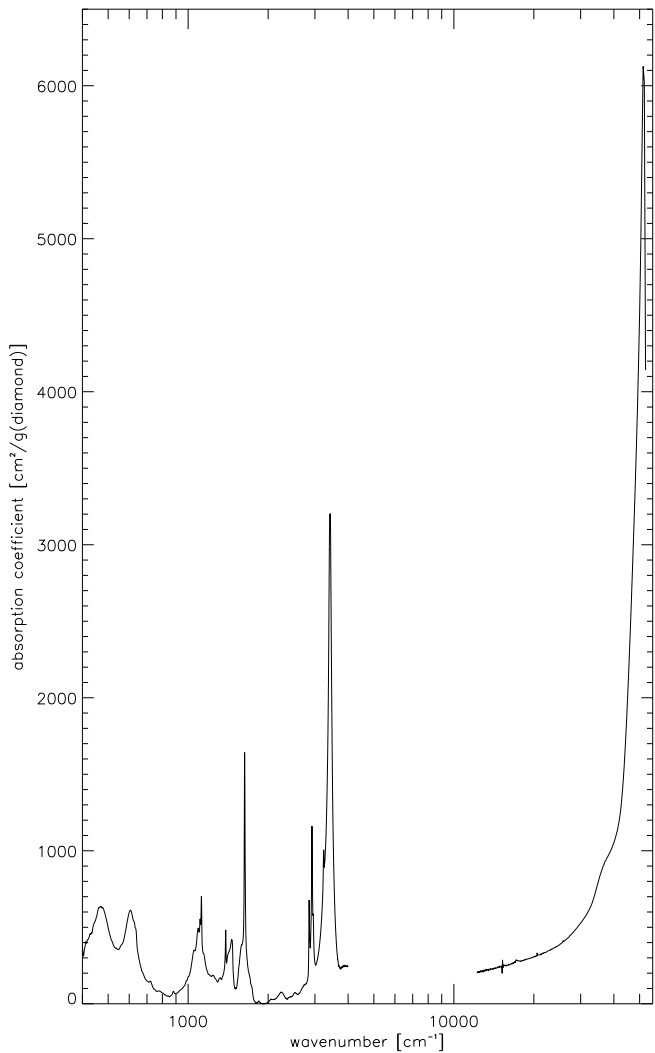}
\caption[]{The monochromatic absorption coefficient
for the presolar diamonds, as derived from the IR and UV/VIS
absorbance spectra.}
\end{figure}

With the measured $I/I_{0}$, being dimensionless, and $\rho$ and $l$
given in the units described above,  the  monochromatic absorption
for the
presolar diamonds comes out in units of cm$^{2}$ per gram of
diamonds which is shown in
Fig.\,5. We notice that with the correction for scattering, the value of the
absorption coefficient at the upper wavenumber of our infrared
measurements is almost identical to the value at the low-wavenumber end of the
UV/VIS measurements. Further, Mutschke et al. (1995) notice that this
spectral region in their measurements of the presolar diamonds from
Murchison, is featureless, which is a well known characteristic of 
terrestrial diamonds, too.  A good approximation of the absorption coefficient
in the region from 4000 cm$^{-1}$ to 12200 cm$^{-1}$ is therefore to
approximate it with a constant value, $\kappa$ = 220 cm$^{2}$/g, which we
have also done in the computations of the stellar models and synthetic
stellar spectra presented in the next chapter. 
The full set of data is obtainable on anonymous ftp\footnote{Use userid
anonymous, type your e-mail address as password, type then "cd pub/scan".
You now find the data in the file "diamonds.dat", and a description of how to
use them in the file "diamonds.tex".} from {\sl stella.nbi.dk}.

\subsection{The uncertainty on $\kappa$}

The uncertainty of the monochromatic absorption coefficient, $\kappa$,
as determined from Eq.\,(\ref{UV}) and (\ref{IR}) includes the uncertainty in;
(1) the estimate of the mass, (2) the assumption that the grains are
homogeneously distributed in the sample, and (3) the uncertainty in the
measurement of I/I$_{0}$.  Of these three we estimate, that the
uncertainty in $m_{\rm dia}$ is the dominant factor.
The mass of our presolar diamond
sample for the UV measurements was determined to be 350 $\pm$ 25 $\mu$g, 
giving an uncertainty of 7.1\% and for the IR measurement 
to be 300 $\pm$ 45 $\mu$g implying an uncertainly of 15\% . 
The overall uncertainty on $\kappa$ as determined from Eq.\,(\ref{UV}) and
(\ref{IR}) is therefore around 15\%.

\section{Synthetic stellar spectra}

The knowledge of the monochromatic absorption coefficient for the
nano-diamonds allows construction of self-consistent models of the
stellar environments in which the diamonds can have formed.
We present here self-consistent photospheres and synthetic spectra of
carbon stars with simplified diamond condensation.  We assume that
the diamonds form in the temperature range from 1600\,K to 
1500\,K in such
a way that the amount of \c2h2\ which transforms into diamond grains
increases linearly from 0\% at 1600\,K to a value at 1500\,K which we have
set to 0\%, 5\%, and 10\%, respectively, in various models.
Models where we induce considerably larger amounts of dust
condensation have convergence problems, possibly due to conflicts
with our assumption of hydrostatic equilibrium.
For temperatures below 1500\,K, the condensed fraction of \c2h2 is
assumed to be constant (0\%, 5\%, and 10\%, respectively).
C$_{2}$H$_{2}$ has been suggested as the primary species for diamond
growth by Sharp \& Wasserburg (1993) and Kr{\"u}ger et al.\ (1996),
based on full chemical pathway calculations.
The condensation temperatures we have adopted are
based on the calculation by Sharp \& Wasserburg (1993) and by Lodders \&
Fegley (1995); both of these groups have performed molecular
equilibrium calculations to establish the stability fields of C(s),
TiC(s) and SiC(s) and other high temperature phases under conditions
of different pressures and C/O ratios.

Our computed model photospheres are based on an improved version
(J{\o}rgensen et al.\ 1992) of the {\sc marcs}
code (Gustafsson et al. 1975).  It assumes
hydrostatic equilibrium and local thermodynamic equilibrium (LTE), but
includes effects of sphericity and uses an opacity sampling (OS)
treatment of molecular opacities from approximately 60 million spectral
lines (J{\o}rgensen 1994).
Such models have proven to reproduce well the observed spectral features
of carbon stars (Lambert et al.\ 1986; J{\o}rgensen 1989;
J{\o}rgensen \& Johnson 1991).
Full line opacities of CO, C$_{2}$, CN, CH, \c2h2, C$_{3}$, HCN,
and presolar diamonds were included.
The model calculations were performed with the following parameters:
C/O = 1.17, 1.35, 2.00;
T$_{\rm eff}$ = 2600\,K, 2700\,K, 2800\,K; log(g) = 0.0, Z = Z$_{\odot}$,
and M=M$_{\odot}$.

\begin{figure}
\centering
\leavevmode
\epsfxsize=1.05
\columnwidth
\epsfbox{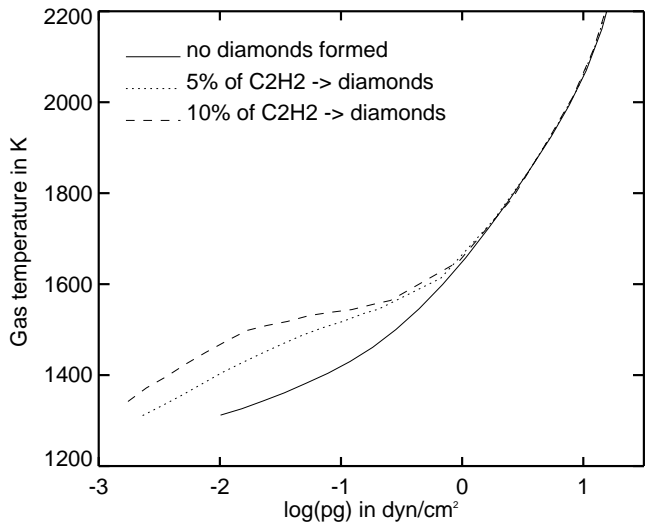}
\caption[]{The gas temperature versus total gas pressure for three carbon
star model atmospheres with \teff\ = 2600\,K, log(g) = 0, C/O = 1.17,
M = M\sun, and Z = Z\sun. The full drawn line represent a model where
no dust is allowed to form. The two other curves
represent models where \c2h2\ gradually (linearly in temperature)
solidifies to diamonds in the temperature range from 1600\,K to 1500\,K.
The maximum degrees of completeness (for T$\le$1500\,K) of the
\c2h2\ consumptions are 5\% and 10\%, for the dotted
and the dashed curves, respectively.  }
\label{models}
\end{figure}

Fig.\,6 shows the temperature versus
gas pressure structure of a model atmosphere of \teff\ = 2600 K,
log(g) = 0, Z = Z\sun, C/O = 1.17, without diamond
formation, and with 5\%, and with 10\%, respectively,
of the \c2h2\ transformed
into diamond over a region of the atmosphere from 1600\,K to 1500\,K.
The result of the inclusion of the diamond opacity into the model
calculation, is a heating of the upper layers of the photosphere
for a given value of the gas pressure. This is a consequence of the 
larger absorption coefficient of diamond than of \c2h2\ in the 
wavelength region just long-ward of the Planck maximum (at 1 $-$ 2 $\mu$m).
As a result of this heating the model with diamond included
produce a slight increase (compared to the model without diamond)
in the spectral
flux of the central part of the 3 $\mu$m (\c2h2) band as well as of the 
spectral region of the diamond features themselves.
The effect is, however, too small to
make a clear identification of possible stellar extrasolar diamonds
likely on this basis. It should also be mentioned that the effect of
reducing the intensity of the 3 $\mu$m spectral feature due to the
photospheric heating, is smaller than the 
effect caused by reducing the adopted gravity of the model atmosphere
within observationally determined limits (J{\o}rgensen 1989), so
this indirect measure of the diamond absorption is not applicable either.

\begin{figure}
\centering
\leavevmode
\epsfxsize=1.05
\columnwidth
\epsfbox{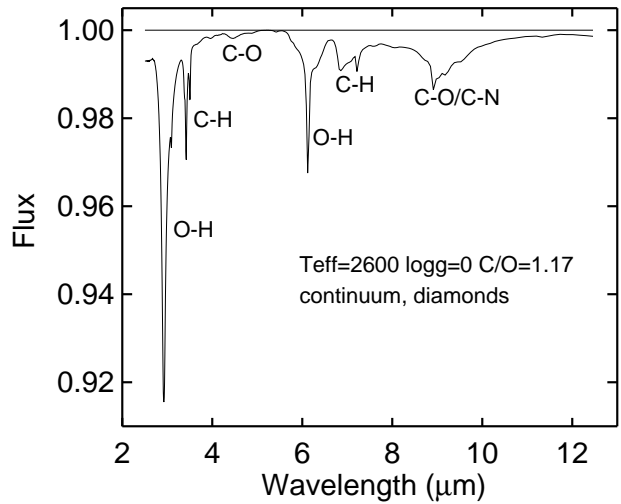}
\caption[]{The absorption spectrum from 2 to 12 $\mu$m (5,000 to 800
cm$^{-1}$) of presolar diamonds in 
a carbon star atmosphere (model as in Fig.\ 6, with 10\% condensation),
calculated based on the measured presolar diamond absorption coefficient 
shown in Fig.\ 5. Our assignments (from Table 1) are indicated
along with the spectrum.}
\label{diam_spc}
\end{figure}

\begin{figure}
\centering
\leavevmode
\epsfxsize=1.05
\columnwidth
\epsfbox{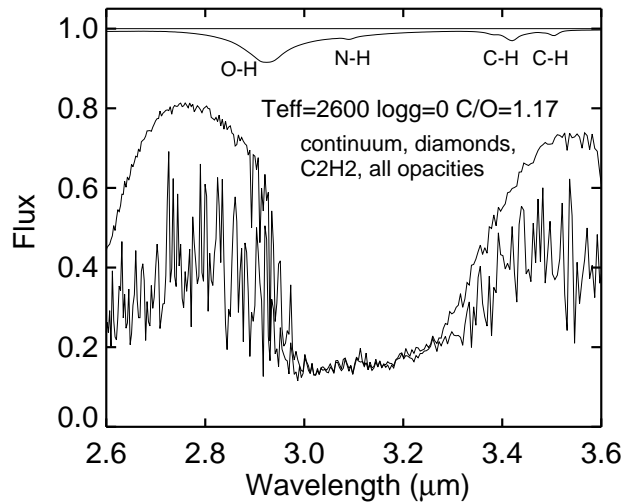}
\caption[]{Same as Fig.\ 7 (although limited to the spectral region 
2.6 $\mu$m $-$ 3.6 $\mu$m). In addition to the continuum and the
diamond spectrum, also the \c2h2\ spectrum, and the full self-consistent
synthetic spectrum based on all included opacity sources, are shown.  }
\label{total_spc}
\end{figure}

Fig.\,7 shows the diamond
spectrum, normalised to the true continuum, for the model from Fig.\,6
where 10\% of the \c2h2\ has transformed into diamonds.
This spectrum is essentially the monochromatic absorption coefficient
modified by the radiation field and the partial pressures through the
atmosphere. The same features as in Fig.\,3 are therefore
seen here too. Some of these may be artifacts caused by the extraction
procedure, as discussed above, but all the measured spectral
features are shown here for completeness.

It is seen that the strongest spectral feature (at
$\approx$ 3 $\mu$m) from our
measurements of the presolar diamonds gives rise to a reduction of the
model spectral flux of 8\%.  In Fig.\,8 is seen that the left 
over \c2h2\ in the same model gives rise to a
flux reduction of between 20\% and 80\% in the 3 $\mu$m spectral region.
When also the other
molecules in this region are taken into account, the outcome 
of synthetic spectrum computations based on a fixed model structure,
is almost identical spectra independent of whether the diamonds
are included in the spectra or not.
In addition, the 3 $\mu$m diamond feature is likely to be
an artifact of the chemical extraction procedure, as discussed
above, so that the real intrinsic
diamond spectral features may be even weaker. Also at 6 and 9 $\mu$m,
where the total molecular absorption is weaker than at 3 $\mu$m, the 
diamond features are weak compared to the molecular spectral features.

Generally, grain nucleation (as opposed to grain growth)
is the bottle neck in stellar dust formation.
Polycyclic Aromatic Hydrocarbons (PAH)
are often assumed to be the molecular nucleation seeds for dust growth
in carbon stars. In hydrodynamic model atmospheres, however,
the time scales for PAH formation are too long compared to the
dynamical time scales, and in the corresponding
hydrostatic photospheres the gas temperature is too high for PAH
formation (Helling et al.\ 1996). 
The relatively modest opacity, and higher condensation temperature
of the diamonds than that of the PAH molecules, may cause
nucleation of diamond grains at relatively high atmospheric densities,
where the velocity field is still negligible (hence, the hydrostatic
approximation is acceptable here). If nucleation of diamond
dust (as oppose to PAH) can act as seeds for grain growth in carbon
stars, the long dynamical time scales in the region of diamond
condensation may therefore contribute to 
the solution of the nucleation problem mentioned above.

If diamonds are the nucleation seeds of other dust grains,
then we should of course expect that the presolar diamonds
were part of agglomerates or
part of heterogeneous larger grains when they left the
parent stellar atmosphere, rather than being pure diamonds
(but our considerations above about the photospheric spectra would still
apply, since the small, 
high-temperature diamond seeds will form as individual grains
in the pseudo-hydrostatic, denser photosphere before the agglomeration).
Such larger grains may, however, have been destroyed in
the interstellar space or in the solar nebula. If they survived
all the way to being included in the carbonaceous chondrites,
they would unfortunately most likely be dissolved during the present
chemical extraction procedure, and the development of a more gentle
extraction procedure could therefore add important
new information about the origin of the presolar diamonds.

\section{Discussion and Conclusion}

The obtained spectra of the presolar diamonds show many features which are
mostly dominated by O--H, C--H and C--N features.  
The O--H features does probably not
originate from when the nano-diamonds were formed in a carbon/hydrogen rich
environment, but is rather an artifact originating from the very rough 
chemical treatment that is used to extract the diamonds, but the C--H, C--N
and N--H (if the nitrogen inclusions are situated at the surface) bonds could
very well be responsible for the features we should 
expect to detect if diamonds are present
in a stellar environment.

This mean that the UV/VIS features around 217 and 270 nm (paired N in
diamond), the feature around 3200 cm$^{-1}$ (N--H), the features around
2900 cm$^{-1}$ (C--H) and 1100 cm$^{-1}$ (C--N/C--C/interstitial N) and
the features around 350 cm$^{-1}$ (C--N--C) and 125 cm$^{-1}$ (unidentified) are
all promising features when trying to establish the presence of nano-diamonds.

Knowledge of the monochromatic absorption coefficient is necessary in order to
include grains in model atmosphere calculations and in synthetic spectrum
calculations, which are needed for comparison with high-resolution observed
spectra.  In the models presented here the
diamond spectrum is weak compared to the molecular spectral
features, indicating that an observational identification of diamonds
not will be straight forward. 
The effect on the atmospheric structure is however substantial, and
nano-diamonds could play an important role in the grain nucleation process.

We have presented here the computed spectrum as it would look like if the
diamonds form in carbon stars (under the simplified assumptions described
above).  Our measured monochromatic absorption coefficient (which is available
from the authors) can likewise be used by other authors, doing models of
other objects, for predicting the spectrum from such objects.

Beside carbon stars, the other possible  
candidates for the diamond formation includes 
novae (Clayton et al. 1995), 
carbon rich Wolf-Rayet stars (Tielens 1990; Arnould et al.\ 1993), and
young expanding supernova remnants with dust originating from the deeper
layers of the star (Clayton 1989; Clayton et al. 1995).

Until now, the observational searches for presolar diamond
grains have focused on the interstellar medium.
One of the most encouraging results is the present identification by
Allamandola et al.\ (1992) of an
absorption band in four dense molecular clouds (proto stars) at 3.47 $\mu$m
(2882 cm$^{-1}$),
which they attributed to a tertiary C--H stretching mode, and
tentatively interpreted as due to hydrogenate nano-diamonds. 
We notice that this absorption
band is in the exact right region to be a mix of the two
absorption features we observe at 
2927 cm$^{-1}$ and 2856 cm$^{-1}$, present 
in our infrared spectrum of the presolar diamonds.  
Other searches for diamonds in interstellar space have, 
however, mainly reached negative results (Sandford et al.\ 1991 and references
therein), 
and for this there might
be at least three reasons;
one is that the presence of functional groups 
attached to the surface of the grains will alter the
absorption features.  The second reason is that the nano-diamonds might be
included in other dust grains, if they work as the nucleation seed, the
third reason is that the stellar grains are likely to 
be covered with ices while in interstellar space.  
The possible interstellar diamond spectrum is therefore expected to be quite
different (or absent if the grain mantle or ice cover is sufficiently thick) 
from the
obtained laboratory spectrum. 

Diamonds is the major known presolar component in meteorites, but their
origin is still a puzzle.  We have presented here data, and preliminary 
analysis based on these data, which might make it possible to identify
their place of origin.
An observational identification of the stellar source of the presolar
grains would lead to improved understanding of the upper layers of stellar
atmospheres, grain formation, the mass loss process, and of the
detailed chemical evolution of our Galaxy.

\begin{acknowledgements}            

The authors would like to thank Roy Lewis for detailed advice on the
extraction of the presolar grains, Carsten N{\o}rg{\aa}rd for making the
CVD diamonds, Merete Torpe for help with the UV/VIS spectrometer, Lykke
Ryelund for help with the IR-spectrometer and Daniel H{\o}jgaard
Christensen for valuable discussions.
We thank the refeere E.\,Sedlmayr for valuable comments on the
manuscript.

\end{acknowledgements}

\end{document}